\documentclass[pra, two column, show pacs]{revtex4}%
\usepackage{amsfonts}
\usepackage{amsmath}
\usepackage{amssymb}
\usepackage{graphicx}%
\providecommand{\U}[1]{\protect\rule{.1in}{.1in}}
\begin{document}
\title{Bloch oscillations of spin-orbit-coupled cold atoms in an optical lattice and
spin current generation}
\author{Wei Ji$^{1}$, Keye Zhang$^{1,3}$, Weiping Zhang$^{2,3}$, Lu Zhou$^{1,3}%
$\footnote{Corresponding author: lzhou@phy.ecnu.edu.cn }}
\affiliation{$^{1}$State Key laboratory of Precision Spectroscopy, Quantum Institute of
Light and Atoms, School of Physics and Material Science, East China Normal
University, Shanghai 200241, China}
\affiliation{$^{2}$Department of Physics and Astronomy, Shanghai Jiaotong University and
Tsung-Dao Lee Institute, Shanghai 200240, China}
\affiliation{$^{3}$Collaborative Innovation Center of Extreme Optics, Shanxi University,
Taiyuan, Shanxi 030006, China}

\begin{abstract}
We study the Bloch oscillation dynamics of a spin-orbit-coupled cold atomic
gas trapped inside a one-dimensioanl optical lattice. The eigenspectra of the
system is identified as two interpenetrating Wannier-Stark ladder. Based on
that, we carefully analyzed the Bloch oscillation dynamics and found out that
intraladder coupling between neighboring rungs of Wannier-Stark ladder give
rise to ordinary Bloch oscillation while interladder coupling lead to small
amplitude high frequency oscillation superimposed on it. Specifically
spin-orbit interaction breaks Galilean invariance, which can be reflected by
out-of-phase oscillation of the two spin components in the accelerated frame.
The possibility of generating spin current in this system are also explored.

\end{abstract}

\maketitle

\section{introduction}

\label{sec_introduction}

Bloch oscillation describe that inside a lattice potential a particle will
perform periodic oscillation instead of constant acceleration when subject to
a constant external force. It was first proposed in electronic system
\cite{blochoriginal}, however have not been observed until the use of
semiconductor superlattice \cite{blochexperiment} due to the small lattice
constant and imperfections in conventional crystal. The frequency of Bloch
oscillation is propotional to the applied force $F$, which can have potential
application in precision measurement. Besides that, the dynamics concerning
particles moving in periodic structures is itself important due to that it is
a pure quantum effect and reflects the properties of energy band such as the
topology \cite{topology}. These extends people's interest in Bloch oscillation
beyond the electronic system. Bloch oscillation have been experimentally
observed in optical system \cite{opticalBloch} and ultracold atoms trapped in
an optical lattice \cite{castin1996,niuPRL1996,atomBlochreview}. Recently it
was demonstrated that impurity moving in quantum liquids can also display the
behavior of Bloch oscillation \cite{impurity,impuritytheory}. Theoretically
Bloch oscillation can be well-understood within adiabatical approximation in
which the particles move in Bloch energy band under the action of the force
\cite{castin1996}. The eigenstate of Bloch oscillation is also well-known as
Wannier-Stark ladder (WSL) \cite{WSLreview}.

On the other hand besides the external centre-of-mass motion, particles
possess internal degree-of-freedom such as the electronic spin. Pseudospin can
also be constructed from the atomic internal energy level structure. Through
the mechanism of spin-orbit (SO) coupling particle's orbital motion can be
connected to its spin dynamics and lead to rich physics. Recently SO coupling
have been successfully implemented in neutral atom
\cite{soReview,1dso,chenPRL2012}. Along with that, interesting physics have
been predicted in SO-coupled atomic system such as dipole oscillation
\cite{chenPRL2012,puPRA2012}, Zitterbewegung \cite{leblancNJP2013,quPRA2013R},
spin-dependent pairing \cite{dongPRA2013}, SO-modulated Anderson localization
\cite{zhouPRA2013,orsoPRL2017,shermanPRL2015}, SO-modulated atom optics
\cite{zhouatomoptics} and exotic dynamics \cite{lanPRA2014,EngelsPRL2015,flat
band,panPRA2016,engelsPRL2017}.

Then it is natural to ask how Bloch oscillation will be affected by SO
interaction. In the present work we will investigate the Bloch oscillation
dynamics of SO-coupled cold atoms in a one-dimensional optical lattice. An
important motivation lies in the recent achievement of SO-coupled
Bose-Einstein condendates (BEC) in a one-dimensional optical lattice
\cite{EngelsPRL2015}, which guarantee that the results obtained here can be
readily observed in experiment. In previous theoretical works, Larson and
co-workers investigated Bloch oscillation of SO-coupled BEC in a
two-dimensional optical lattice, in which transverse spin current and atomic
Zitterbewegung are predicted \cite{LarsonPRA2010}. Bloch oscillation of a
SO-coupled helicoidal molecule was studied by Caeteno in \cite{caetanoPRB2014}%
. Kartashov \textit{et al}. studied Bloch oscillation in one-dimensional
optical and Zeeman lattices in the presence of SO coupling, in which they give
a detailed discussion on the amplitude and wavepacket width of Bloch
oscillation \cite{KartashovPRL2016}. Although the WSL eigen-spectra have been
given in \cite{KartashovPRL2016}, its relation with the oscillation dynamics
was not clarified yet. Here we will solve the dynamics using the theory of
WSL. We show that one can understand the properties of Bloch oscillation
dynamics in the presence of SO coupling via analyzing the coupling of WSLs.
Especially in the case with finite Zeeman detuning which was not considered in
\cite{KartashovPRL2016}, the two spin components will display unusual
out-of-phase oscillation. In addition we show how this can serve as an
unambiguous proof of broken Galilean invariance caused by SO interaction.
Since SO interaction can play a crucial role in generating and manipulating
spin current \cite{SpinCurrentScience}, we'll also look into the possibility
of generating spin current in the present one-dimensional system.

The article is organized as follows: In Sec. \ref{sec_model} we present our
model and the dynamics are solved with WSL. Section \ref{sec_bloch} is devoted
to the detailed discussion of Bloch oscillation. The possibility of generating
spin current in the present system is explored in Sec. \ref{sec_current}.
Finally we conclude in Sec. \ref{sec_summary}.

\section{model}

\label{sec_model}

\begin{figure}[h]
\includegraphics[width=8cm]{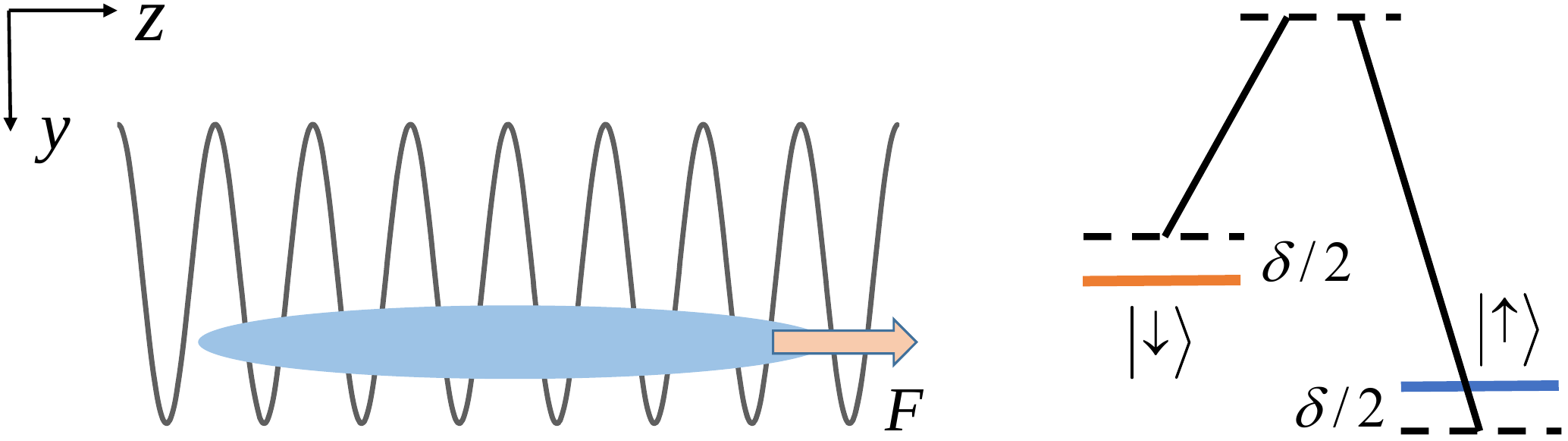}\caption{{\protect\footnotesize (Color
online) Schematic diagram showing the system under consideration.}}%
\label{fig_scheme}%
\end{figure}

As shown in Fig. \ref{fig_scheme}, our model is based on the recent experiment
\cite{EngelsPRL2015} with a $^{87}$Rb BEC prepared in a one-dimensional
optical lattice along the $z$-direction, inside which the effective SO
interaction is induced via coupling the $\left\vert 1,-1\right\rangle $
($\left\vert \downarrow\right\rangle $) and $\left\vert 1,0\right\rangle $
($\left\vert \uparrow\right\rangle $) hyperfine states with Raman lasers. In
addition to that, here we consider that a constant external force $F$ is
exerted on the atoms via tilting the optical lattice. The effective
single-particle Hamiltonian reads%
\begin{align}
\hat{H}  &  =\hat{H}_{SO}+U_{0}\sin^{2}\left(  k_{l}z\right)  -Fz,\nonumber\\
\hat{H}_{SO}  &  =\frac{\left(  p_{z}-\hat{A}\right)  ^{2}}{2m}+\frac
{\hbar\Omega}{2}\hat{\sigma}_{x}+\frac{\hbar\delta}{2}\hat{\sigma}_{z},
\label{eq_single_particle_Hamiltonian}%
\end{align}
in which the SO coupling is embodied in the effective vector potential
$\hat{A}=-m\alpha\hat{\sigma}_{z}$ ($\alpha=\hbar k_{R}/m$ characterizes SO
coupling strength with $k_{R}$ the Raman beam wavevector), $\Omega$ is the
Raman coupling strength with $\delta$ the two-photon detuning. The periodic
potential is characterized by the depth $U_{0}$ and period $d=\pi/k_{l}$.

By performing lowest energy band truncation and assuming tight binding
approximation, Hamiltonian (\ref{eq_single_particle_Hamiltonian}) can be
expanded in the $\sigma$-Wannier basis $\left\vert j,\sigma\right\rangle $
(with $j$ the lattice site index) as%
\begin{align}
\hat{H}  &  =\sum_{j}\left\{  \left[  -\frac{J}{2}\cos\left(  \pi
\gamma\right)  \sum_{\sigma}\left\vert j,\sigma\right\rangle \left\langle
j+1,\sigma\right\vert \right.  \right. \nonumber\\
&  \left.  +i\frac{J}{2}\sin\left(  \pi\gamma\right)  \left(  \left\vert
j,\uparrow\right\rangle \left\langle j+1,\uparrow\right\vert -\left\vert
j,\downarrow\right\rangle \left\langle j+1,\downarrow\right\vert \right)
\right. \nonumber\\
&  \left.  \left.  +\frac{\hbar\Omega}{2}\left\vert j,\uparrow\right\rangle
\left\langle j,\downarrow\right\vert +H.c.\right]  -Fd\sum_{\sigma}j\left\vert
j,\sigma\right\rangle \left\langle j,\sigma\right\vert \right. \nonumber\\
&  \left.  +\frac{\hbar\delta}{2}\left(  \left\vert j,\uparrow\right\rangle
\left\langle j,\uparrow\right\vert -\left\vert j,\downarrow\right\rangle
\left\langle j,\downarrow\right\vert \right)  \right\}
\label{eq_single_band_Hamiltonian}%
\end{align}
in which the spin-dependent hopping matrix element $\hat{T}=J\exp\left(
-i/\hbar\int\hat{A}dl\right)  /2$ is obtained through Peierls substitution
\cite{peierls}, $J$ is the tunneling amplitude without SO coupling,
$\gamma=k_{R}/k_{l}$. $J$ can be calculated as%
\begin{equation}
J=-2\int dzw_{j+1}\left(  z\right)  \left[  -\frac{d^{2}}{dz^{2}}+U_{0}%
\sin^{2}\left(  k_{l}z\right)  \right]  w_{j}\left(  z\right)  ,
\label{eq_hopping_matrix_element}%
\end{equation}
with $w_{j}\left(  z\right)  =w\left(  z-z_{j}\right)  $ is the Wannier state
of the lowest energy band at the $j$-th site which can be obtained numerically
\cite{lattice parameter}. Here we consider the case of $U_{0}>0$ with
$z_{j}=jd$.

\begin{figure}[h]
\includegraphics[width=8cm]{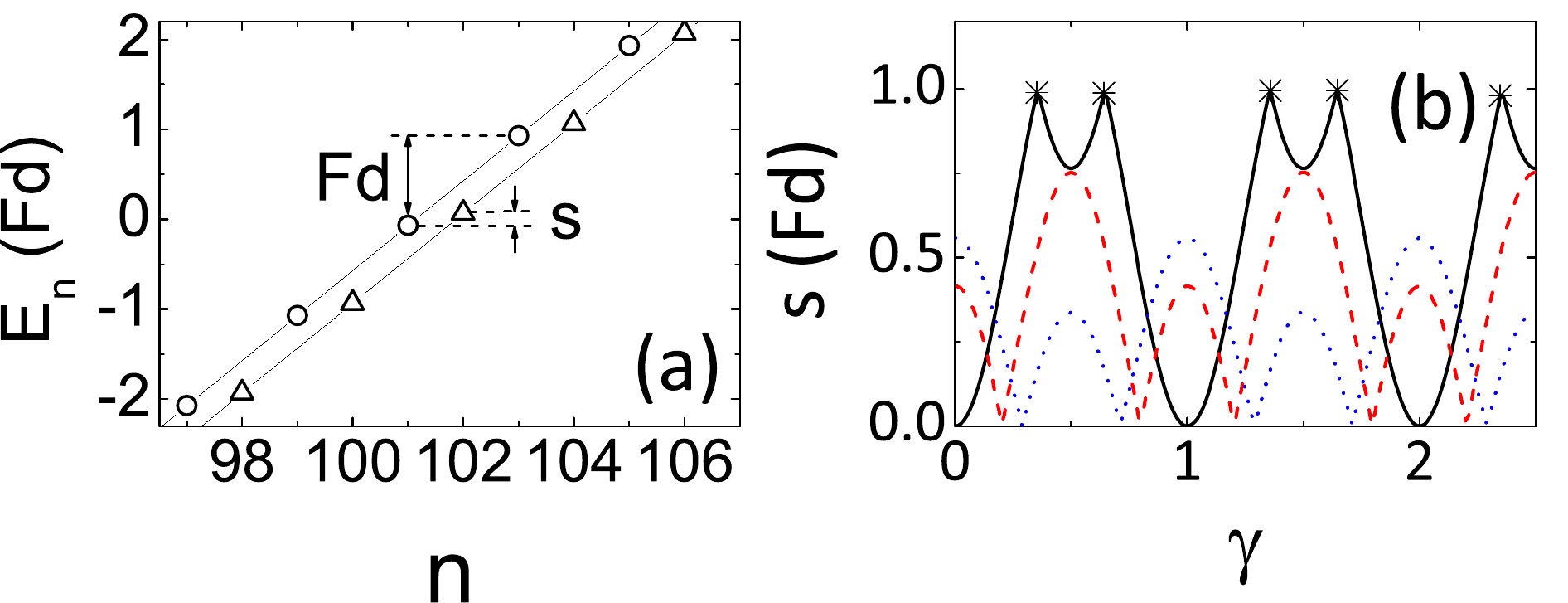}\caption{{\protect\footnotesize (Color
online) (a) Eigenenergy spectra of the system under consideration. The spectra
consists of two interpenetrating WSL, the intraladder spacing of both ladder
is }$Fd$ {\protect\footnotesize while the interladder spacing is }%
$s${\protect\footnotesize . (b) }$s$ {\protect\footnotesize versus }$\gamma$
{\protect\footnotesize at }$\delta=0$ {\protect\footnotesize (black solid
line), }$\delta=0.2\Omega$ {\protect\footnotesize (red dashed line) and
}$\delta=0.5\Omega$ {\protect\footnotesize (blue dotted line). The asterisks
mark the values of }$\gamma$ {\protect\footnotesize at which }$s=Fd$%
{\protect\footnotesize . The other parameters are set as }$J=10Fd$
{\protect\footnotesize and }$\hbar\Omega=80Fd${\protect\footnotesize .}}%
\label{fig_wsl}%
\end{figure}

In order to find out the eigenstates of Hamiltonian
(\ref{eq_single_band_Hamiltonian}), it will be more convienent to transform it
into the Bloch basis via the Fourier transformation \cite{HartmannNPJ2004}%
\begin{equation}
\left\vert q,\sigma\right\rangle =\sqrt{\frac{d}{2\pi}}\sum_{j=-\infty
}^{\infty}\left\vert j,\sigma\right\rangle e^{iqjd}. \label{eq_bloch_basis}%
\end{equation}
One can then obtain%
\begin{equation}
\hat{H}\left(  q\right)  =\left\langle q\left\vert \hat{H}\right\vert
q\right\rangle =\left(
\begin{array}
[c]{cc}%
H_{d}^{+} & \hbar\Omega/2\\
\hbar\Omega/2 & H_{d}^{-}%
\end{array}
\right)  , \label{eq_bloch_Hamiltonian}%
\end{equation}
with $H_{d}^{\pm}=-J\cos\left(  qd\mp\pi\gamma\right)  \pm\hbar\delta
/2-iF\partial/\partial q$. The eigenvalue problem then resort to
\begin{subequations}
\label{eq_eigenvalue}%
\begin{gather}
-iF\frac{\partial\psi_{\uparrow}\left(  q\right)  }{\partial q}-J\cos\left(
qd-\pi\gamma\right)  \psi_{\uparrow}\left(  q\right)  +\frac{\hbar\delta}%
{2}\psi_{\uparrow}\left(  q\right) \nonumber\\
+\frac{\hbar\Omega}{2}\psi_{\downarrow}\left(  q\right)  =E\psi_{\uparrow
}\left(  q\right)  ,\label{eq_eigenvalue_a}\\
-iF\frac{\partial\psi_{\downarrow}\left(  q\right)  }{\partial q}-J\cos\left(
qd+\pi\gamma\right)  \psi_{\downarrow}\left(  q\right)  -\frac{\hbar\delta}%
{2}\psi_{\downarrow}\left(  q\right) \nonumber\\
+\frac{\hbar\Omega}{2}\psi_{\uparrow}\left(  q\right)  =E\psi_{\downarrow
}\left(  q\right)  , \label{eq_eigenvalue_b}%
\end{gather}
where $\psi\left(  q\right)  =\left[  \psi_{\uparrow}\left(  q\right)
,\psi_{\downarrow}\left(  q\right)  \right]  ^{T}$ is the eigenvector.

Consider that $\psi^{\nu}\left(  q\right)  =\left[  \psi_{\uparrow}^{\nu
}\left(  q\right)  ,\psi_{\downarrow}^{\nu}\left(  q\right)  \right]  ^{T}$ to
be the $\nu$-th eigensolution of Eqns. (\ref{eq_eigenvalue}) with the
corresponding eigenvalue $E_{\nu}$, it can be solved via performing the
Fourier expansion%
\end{subequations}
\begin{align}
\psi_{\uparrow}^{\nu}\left(  q\right)   &  =\sqrt{\frac{d}{2\pi}}\sum
_{m=-M}^{M}A_{m}^{\nu}\exp\left[  iqmd+i\frac{J}{Fd}\sin\left(  qd-\pi
\gamma\right)  \right]  ,\nonumber\\
\psi_{\downarrow}^{\nu}\left(  q\right)   &  =\sqrt{\frac{d}{2\pi}}\sum
_{m=-M}^{M}B_{m}^{\nu}\exp\left[  iqmd+i\frac{J}{Fd}\sin\left(  qd+\pi
\gamma\right)  \right]  , \label{eq_fourier expansion}%
\end{align}
where $A_{m}^{\nu}$ and $B_{m}^{\nu}$ are expansion coefficients with the
truncation number $M$. Through numerical calculation we found that $M=50$ to
be a good approximation for the parameters considered in the present work.
Substitute (\ref{eq_fourier expansion}) into Eqns. (\ref{eq_eigenvalue}), one
can have%
\begin{gather}
\frac{\hbar\Omega}{2}\sum_{m^{\prime}}i^{m-m^{\prime}}J_{m-m^{\prime}}\left(
\frac{2J}{Fd}\sin\left(  \pi\gamma\right)  \right)  B_{m^{\prime}}^{\nu
}\nonumber\\
+\left(  mFd+\frac{\hbar\delta}{2}\right)  A_{m}^{\nu}=E_{\nu}A_{m}^{\nu
},\nonumber\\
\frac{\hbar\Omega}{2}\sum_{m^{\prime}}\left(  -i\right)  ^{m-m^{\prime}%
}J_{m-m^{\prime}}\left(  \frac{2J}{Fd}\sin\left(  \pi\gamma\right)  \right)
A_{m^{\prime}}^{\nu}\nonumber\\
+\left(  mFd-\frac{\hbar\delta}{2}\right)  B_{m}^{\nu}=E_{\nu}B_{m}^{\nu},
\label{eq_coupled series}%
\end{gather}
with $J_{n}\left(  z\right)  $ the $n$th-order Bessel functions of the first
kind. One can then numerically solve Eqns. (\ref{eq_coupled series}) and
obtain the coefficients $A_{m}^{\nu}$, $B_{m}^{\nu}$ and the corresponding
eigenenergy $E_{\nu}$. The Wannier amplitudes of the corresponding eigenvector
read%
\begin{align}
W_{j,\uparrow}^{\nu}  &  =\sum_{m}A_{m}^{\nu}J_{-j-m}\left(  \frac{J}%
{Fd}\right)  e^{i\left(  j+m\right)  \pi\gamma},\nonumber\\
W_{j,\downarrow}^{\nu}  &  =\sum_{m}B_{m}^{\nu}J_{-j-m}\left(  \frac{J}%
{Fd}\right)  e^{-i\left(  j+m\right)  \pi\gamma}. \label{eq_wannier amplitude}%
\end{align}

\ In the case without SO coupling the eigenenergy of Eqs. (\ref{eq_eigenvalue}%
) is known as WSL \cite{Wannier1960}, which consists of quantized energy
levels with equal energy spacing $Fd$. In the presence of SO coupling WSL
still exists, as can be seen from the Hamiltonian
(\ref{eq_single_particle_Hamiltonian}) with $\hat{H}\left(  z\right)
\psi\left(  z+d\right)  =\left(  E+Fd\right)  \psi\left(  z+d\right)  $.
However the coupling between two pseudo-spin states will lead to two
interpenetrating WSL which positioned symmetrically around $0$
\cite{KartashovPRL2016}, with an intra-ladder separation $s$, as shown in Fig.
\ref{fig_wsl}(a). The inter-ladder spacing within the two WSL is still $Fd$.
By considering that, we can label the WSL eigenenergy with $\nu_{1\left(
2\right)  }$ and $E_{\nu_{1\left(  2\right)  }}=\nu_{1\left(  2\right)  }Fd\mp
s/2$. The intra-ladder spacing $s$ is a composite function of $\gamma$,
$\Omega$ and $\delta$. As shown in Fig. \ref{fig_wsl}(b), $s$ is a periodic
function of $\gamma$. When $\delta=0$, $s=0$ for integer values of $\gamma$,
the two WSL overlaps. This can be seen from that Eqs. \ref{eq_eigenvalue}(a)
and (b) are the same by replacing $\psi_{\uparrow}\left(  q\right)
\rightarrow\psi_{\downarrow}\left(  q\right)  $ at $\delta=0$ and integer
$\gamma$, signaling identical dynamics for the two spin components.
Interestingly in addition to that, at some specific values of $\gamma$ maked
by asterisks in Fig. \ref{fig_wsl}(b) $s=Fd$, also indicating overlaping WSL.
A nonzero $\delta$ separates the two ladder even at $\gamma=0$.

The relation between the WSL spectrum and dynamics can be understood from the
mean velocity. The velocity operator can be defined as $d\hat{z}/dt=i\left[
\hat{H},\hat{z}\right]  /\hbar$, using the Hamiltonian
(\ref{eq_single_band_Hamiltonian}) and assume the atomic wavefunction
$\left\vert \psi\left(  t\right)  \right\rangle =\sum_{j,\sigma}\psi
_{j,\sigma}\left(  t\right)  \left\vert j,\sigma\right\rangle $, one can
calculate the mean velocity as%
\begin{align}
\frac{dz}{dt}  &  =\frac{Jd}{\hbar}\sum_{j}\operatorname{Im}\left[
e^{-i\pi\gamma}\psi_{j,\uparrow}^{\ast}\left(  t\right)  \psi_{j+1,\uparrow
}\left(  t\right)  \right. \nonumber\\
&  \left.  +e^{i\pi\gamma}\psi_{j,\downarrow}^{\ast}\left(  t\right)
\psi_{j+1,\downarrow}\left(  t\right)  \right]  . \label{eq_velocity1}%
\end{align}
Then one can take advantage of Wannier-Stark eigenstates by considering that
$\psi_{j,\sigma}\left(  t\right)  =\sum_{\nu}a_{\nu}W_{j,\sigma}^{\nu}%
\exp\left(  -iE_{\nu}t/\hbar\right)  $ with $a_{\nu}=\sum_{j,\sigma
}W_{j,\sigma}^{\nu\ast}\psi_{j,\sigma}\left(  0\right)  $, and the mean
velocity can be expressed as%
\begin{align}
\frac{dz}{dt}  &  =\frac{Jd}{\hbar}\sum_{\nu,\nu^{\prime}}\operatorname{Im}%
\left\{  a_{\nu}^{\ast}a_{\nu^{\prime}}\left[  \sum_{j}W_{j,\uparrow}^{\nu
\ast}W_{j+1,\uparrow}^{\nu^{\prime}}e^{-i\pi\gamma}\right.  \right.
\nonumber\\
&  \left.  \left.  +\sum_{j}W_{j,\downarrow}^{\nu\ast}W_{j+1,\downarrow}%
^{\nu^{\prime}}e^{i\pi\gamma}\right]  e^{i\left(  E_{\nu}-E_{\nu^{\prime}%
}\right)  t/\hbar}\right\}  . \label{eq_velocity2}%
\end{align}
The particle mean position $z\left(  t\right)  =z_{\uparrow}\left(  t\right)
+z_{\downarrow}\left(  t\right)  $ can then be derived via integrating Eq.
(\ref{eq_velocity2}) over time, in which%
\begin{align}
z_{\uparrow\left(  \downarrow\right)  }\left(  t\right)   &  =\sum_{\nu\neq
\nu^{\prime}}\frac{Jd}{E_{\nu}-E_{\nu^{\prime}}}\operatorname{Re}\left\{
a_{\nu}^{\ast}a_{\nu^{\prime}}\sum_{j}W_{j,\uparrow\left(  \downarrow\right)
}^{\nu\ast}W_{j+1,\uparrow\left(  \downarrow\right)  }^{\nu^{\prime}}e^{\mp
i\pi\gamma}\right. \nonumber\\
&  \left.  \times\left[  1-e^{i\left(  E_{\nu}-E_{\nu^{\prime}}\right)
t/\hbar}\right]  \right\}  +z_{\uparrow\left(  \downarrow\right)  }\left(
0\right)  \label{eq_position}%
\end{align}
symbol the mean position of spin-$\sigma$ component. Eq. (\ref{eq_position})
predict that the oscillation frequencies are ruled by the energy difference
between two Wannier-Stark levels with the amplitude of each frequency
inversely propotional to the energy distance of those Wannier-Stark states and
propotional to the overlap of their wavefunctions.

In the absence of SO coupling it is well-known that the Wannier-Stark
eigenstate $W_{j}^{\nu}$ have the form of Bessel function of the first kind
($J_{\nu+j}\left(  z\right)  $) with $W_{j+1}^{\nu}=W_{j}^{\nu+1}$
\cite{HartmannNPJ2004}, then $\sum_{j}W_{j}^{\nu\ast}W_{j+1}^{\nu^{\prime}%
}=\sum_{j}W_{j}^{\nu\ast}W_{j}^{\nu^{\prime}+1}$ take the value $1$ for
$\nu=\nu^{\prime}+1$ and $0$ otherwise. It indicates that in the oscillation
dynamics each rung of the WSL is only coupled to its neighboring rung with
the Bloch frequency $\omega_{B}=\left(  E_{\nu}-E_{\nu^{\prime}}\right)
/\hbar=Fd/\hbar=2\pi/T_{B}$. One can notice that in the presence of SO
coupling the coupled equations (\ref{eq_coupled series}) indicate two WSL in
which any rung of the ladder is coupled to all the rungs of the other ladder,
which will substantially modify the Bloch oscillation dynamics. This will be
discussed in detail in the subsequent section.

\section{bloch oscillation dynamics}

\label{sec_bloch}

\begin{figure}[h]
\includegraphics[width=8cm]{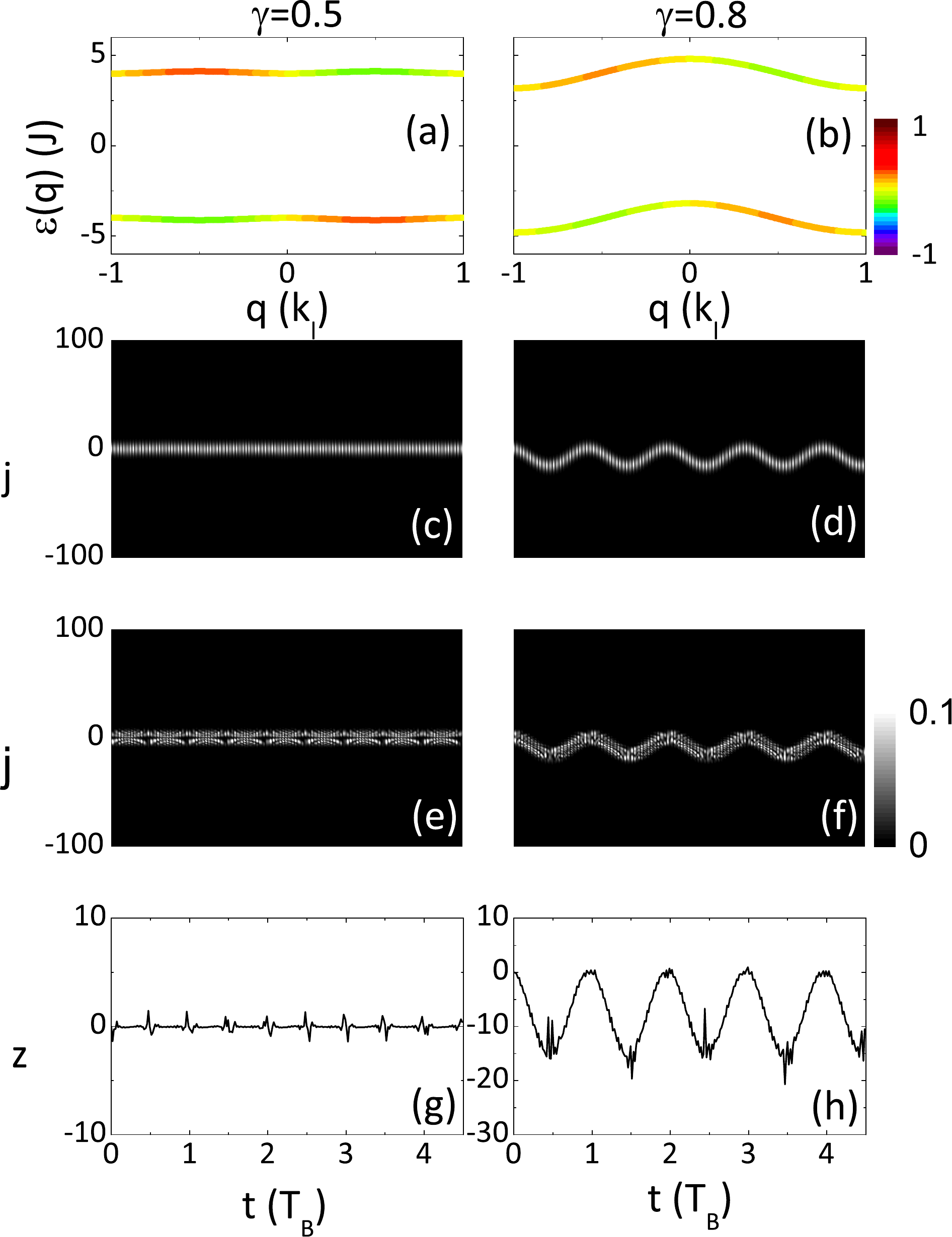}\caption{{\protect\footnotesize (Color
online) (a) and (b) Energy band for an atom in a periodic potential }$U\left(
z\right)  =U_{0}\sin^{2}k_{l}z$ {\protect\footnotesize and subject to SO
interaction, with the color indicating spin polarization }$\left\langle
\hat{\sigma}_{z}\right\rangle ${\protect\footnotesize . (c) and (d) Dynamics
of }$\left\vert \psi_{\uparrow}\right\vert ^{2}${\protect\footnotesize . (e)
and (f) Dynamics of }$\left\vert \psi_{\downarrow}\right\vert ^{2}%
${\protect\footnotesize . (g) and (h) Dynamics of the mean position }%
$z${\protect\footnotesize . The left column correspond to }$\gamma=0.5$
{\protect\footnotesize while the right column correspond to }$\gamma
=0.8${\protect\footnotesize . The other parameters are set as }$J=10Fd$%
{\protect\footnotesize , }$\hbar\Omega=80Fd$ {\protect\footnotesize and
}$\delta=0${\protect\footnotesize .}}%
\label{fig_oscillation}%
\end{figure}

The Bloch oscillation dynamics have been studied in \cite{KartashovPRL2016}
for the case of $\delta=0$. The results predicted there can be well understood
under adiabatical theory. When $F$ is weak enough not to induce interband
transitions the adiabatic approximation can be applied, under which the atoms
move adiabatically along the energy band with the quasimomentum $q\left(
t\right)  =q\left(  0\right)  +Ft/\hbar$. One can predict that the frequency
of Bloch oscillation is propotional to $Fd$ with the amplitude propotional to
the bandwidth. The properties of Bloch oscillation can then be captured via
further looking into the energy band structure, which can be obtained through
diagonalizing the Hamiltonian (\ref{eq_bloch_Hamiltonian}) without the force
($F=0$). This result in a two-band structure with $\varepsilon_{\pm}\left(
q\right)  =-J\cos qd\cos\pi\gamma\pm\sqrt{J^{2}\sin^{2}qd\sin^{2}\pi
\gamma-\hbar\delta J\sin qd\sin\pi\gamma+\hbar^{2}\delta^{2}/4+\hbar^{2}%
\Omega^{2}/4}$. Two major results are predicted in \cite{KartashovPRL2016}:
(i) In analogue to increasing the potential depth $U_{0}$ of the optical
lattice, SO interaction can take the same effect of band flattening \cite{flat
band}. In this case the bloch oscillation amplitude will be suppressed and
thus make it difficult to measure. An example for this is given at
$\gamma=0.5$ with the energy band shown in Fig. \ref{fig_oscillation}(a). (ii)
Since that in the adiabatic approximation the mean velocity of the atom
$v\left(  q\right)  =d\varepsilon\left(  q\right)  /\hbar dq$, the change in
the band structure indicate that the atomic dynamics will subject to strong
modification. As an example, for the band profile at $\gamma=0.8$ shown in
Fig. \ref{fig_oscillation}(b), the initial atomic moving direction will be
reversed.

These phenomena can also be explained using the theory of WSL. By considering
that the eigenstate of the system consists of two interpenetrating WSL, one
can group their contribution to the dynamics into two terms. Similar to the
case without SO coupling, start from Eqs. (\ref{eq_coupled series}) and
(\ref{eq_wannier amplitude}) one can prove that within each ladder
$W_{j+1,\sigma}^{\nu_{i}}=W_{j,\sigma}^{\nu_{i}+1}$ $\left(  i=1,2\text{ label
the two ladders}\right)  $ still hold true, then according to Eq.
(\ref{eq_position}) one can conclude that in the presence of SO interaction
the Bloch oscillation dynamics in general are still dominated by intra-ladder
coupling between neighboring rungs within each ladder, indicating the
oscillation frequency $T_{B}$.\ At $\delta=0$, due to the symmetry between
spin-$\uparrow$ and $\downarrow$ components, we have $\sum_{j}\left\vert
W_{j,\uparrow}^{\nu}\right\vert ^{2}=\sum_{j}\left\vert W_{j,\downarrow}^{\nu
}\right\vert ^{2}=1/2$, then according to Eqs. (\ref{eq_velocity2}) and
(\ref{eq_position}) one can predict that $z\left(  t\right)  =0$ at
$\gamma=0.5$ and $dz/dt<0$ at $\gamma=0.8$ for initial small $t$, indicating
that Bloch oscillation dynamics are substantially modified by SO interaction.

We assume that initially the atomic wavefunction%

\begin{equation}
\psi_{j}\left(  t=0\right)  =\left(  a\sqrt{\pi}\right)  ^{-1/2}e^{-\left(
j-j_{0}\right)  ^{2}/2a^{2}+iq_{0}jd}\binom{1}{0}, \label{eq_initial}%
\end{equation}
to be a spin-polarized Gaussian wave-packet with width $a$, where $j_{0}$ is
the center of the wave-packet while $q_{0}$ denotes the initial quasimomentum.
In our calculations the parameters are chosen as $j_{0}=0$ and $q_{0}=0$. The
dynamics are simulated using the method of eigenstate expansion and the
results are demonstrated in Figs. \ref{fig_oscillation}(c)-(f), from which one
can see that the results of numerical simulation are consistent with the above
theoretical analysis.

Besides intraladder coupling, interladder coupling also contribute to the
oscillation dynamics. We calculate the value of $\sum_{j}W_{j,\sigma}^{\nu
_{1}\ast}W_{j+1,\sigma}^{\nu_{2}}$ and found out that for relatively large
$\left\vert \nu_{1}-\nu_{2}\right\vert $ (approaching $100$) it really
matters. This can be traced to the symmetry within WSL. Eq.
(\ref{eq_coupled series}) indicate that if $\left(  A_{m},B_{m}\right)  $ are
eigensolutions with eigenvalue $E_{\nu}$, then $\left(  -B_{-m}^{\ast}%
,A_{-m}^{\ast}\right)  $ are eigensolutions with eigenvalue $-E_{\nu}$. Due to
the large energy difference of interladder coupling, it will superimpose small
amplitude high frequency oscillation on the dynamics dominated by intraladder coupling.

An interesting case is that at $\gamma=0.5$, since the intraladder coupling
are canceled out, then the dynamics deviating from $z=0$ is the result of
interladder coupling, which is shown in Fig. \ref{fig_oscillation}(g). One can
observe small amplitude high frequency oscillations, which become prominent
around $t=nT_{B}/2$. Similar behavior can also be observed for $\gamma=0.8$ in
Fig. \ref{fig_oscillation}(h), in which the small oscillations are
superimposed on the traditional Bloch oscillation.\begin{figure}[h]
\includegraphics[width=8cm]{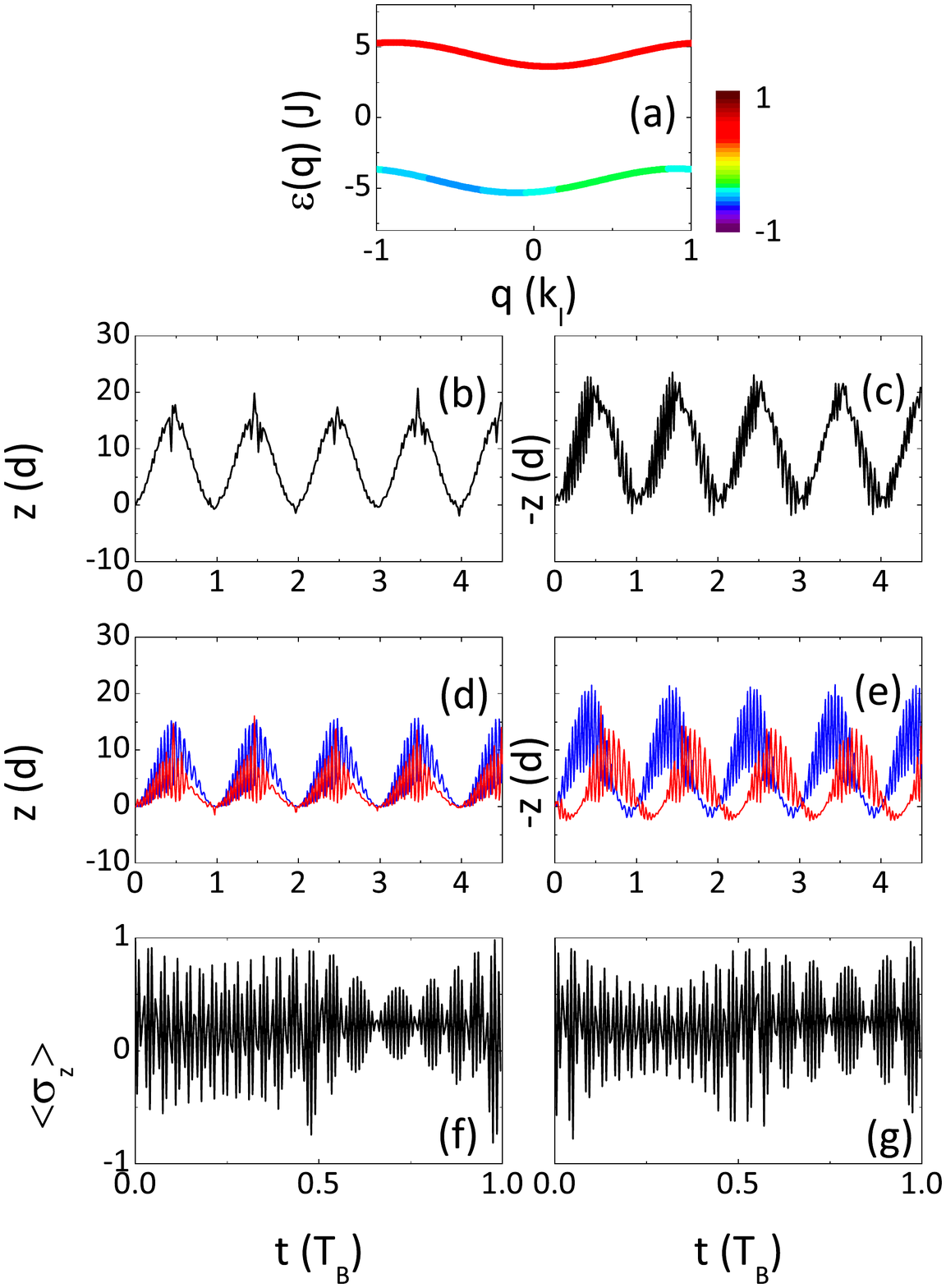}\caption{{\protect\footnotesize (Color
online) (a) Asymmetric energy band at }$\delta=0.5\Omega$
{\protect\footnotesize with the color indicating spin polarization
}$\left\langle \hat{\sigma}_{z}\right\rangle ${\protect\footnotesize . (b)
Dynamics of mean position }$z$ {\protect\footnotesize with the exerting force
}$F$ {\protect\footnotesize along the }$+z$ {\protect\footnotesize direction.
Same dynamics of }$z_{\uparrow}$ {\protect\footnotesize (blue line) and
}$z_{\downarrow}$ {\protect\footnotesize (red line) are shown in (d). (c) and
(e) Same as (b) and (d) except that the force }$F$ {\protect\footnotesize is
exerted along the }$-z$ {\protect\footnotesize direction. (f) and (g) Mean
value of pseudo-spin }$\left\langle \hat{\sigma}_{z}\right\rangle $
{\protect\footnotesize versus time for the force }$F$
{\protect\footnotesize exerted along }$+z$ {\protect\footnotesize and }$-z$
{\protect\footnotesize direction, respectively. The other parameters are set
as }$\gamma=0.2${\protect\footnotesize , }$J=10Fd$ {\protect\footnotesize and
}$\hbar\Omega=80Fd${\protect\footnotesize .}}%
\label{fig_finite delta}%
\end{figure}

The Klein four-group \cite{KartashovPRL2016} or $\mathcal{CPT}$ symmetry
\cite{symmetry} is conserved by the Hamiltonian $\hat{H}_{SO}+U_{0}\sin
^{2}\left(  k_{l}z\right)  $ at $\delta=0$, then in the corresponding energy
band the eigenfunctions are symmetric for spin-$\uparrow$ and $\downarrow$
($\psi_{\uparrow}\left(  q\right)  =\psi_{\downarrow}\left(  -q\right)  $)\ at
the centre and edge of Brillouin zone, which can also be seen from Eqs.
(\ref{eq_eigenvalue}). Then within adiabatical theory one can predict that
$\left\langle \hat{\sigma}_{z}\right\rangle =0$ when the atoms pass through
the centre and edge of Brillouin zone. However this symmetry is broken at
finite $\delta$. At finite $\delta$ the upper energy band and the lower one
are shifted to opposite directions with respect to $q=0$, as shown in Fig.
\ref{fig_finite delta}(a). Physically this band asymmetry can be captured
through Bloch oscillation via exerting force in opposite directions. The
numerical results are shown in Figs. \ref{fig_finite delta}(b) and (c), in
which a force $F$ are considered to be exerted along the $+z$ and $-z$
direction, respectively. At $\delta=0$ one would expect that these two
dynamics are identical, here the different dynamics signal the energy band
asymmetry. Since the atomic initial state can be viewed as the superposition
of the upper and lower eigenstate of the two bands, then in adiabatic limit
they will subject to different dispersion under the action of the force. This
cannot take place at $\delta=0$ where the energy band are always symmetric and
the two bands possess almost identical dispersion. The combined effect will
lead to different oscillation dynamics for the two spin components as we
illustrated in Fig. \ref{fig_finite delta}(e), the dynamics become
out-of-phase for the two spin components. One can also notice that in Fig.
\ref{fig_finite delta}(d) the high frequency oscillations for the two
components are out-of-phase, this is because $W_{j,\uparrow}^{\nu\ast
}W_{j+1,\uparrow}^{\nu^{\prime}}=-W_{j,\downarrow}^{\nu\ast}W_{j+1,\downarrow
}^{\nu^{\prime}}$ for interladder couplings. In the meanwhile, $\left\langle
\hat{\sigma}_{z}\right\rangle $ deviate from $0$ when the wavepacket passes
through the centre and edge of the Brillouin zone, as shown in Figs.
\ref{fig_finite delta}(f) and (g).

\begin{figure}[h]
\includegraphics[width=8cm]{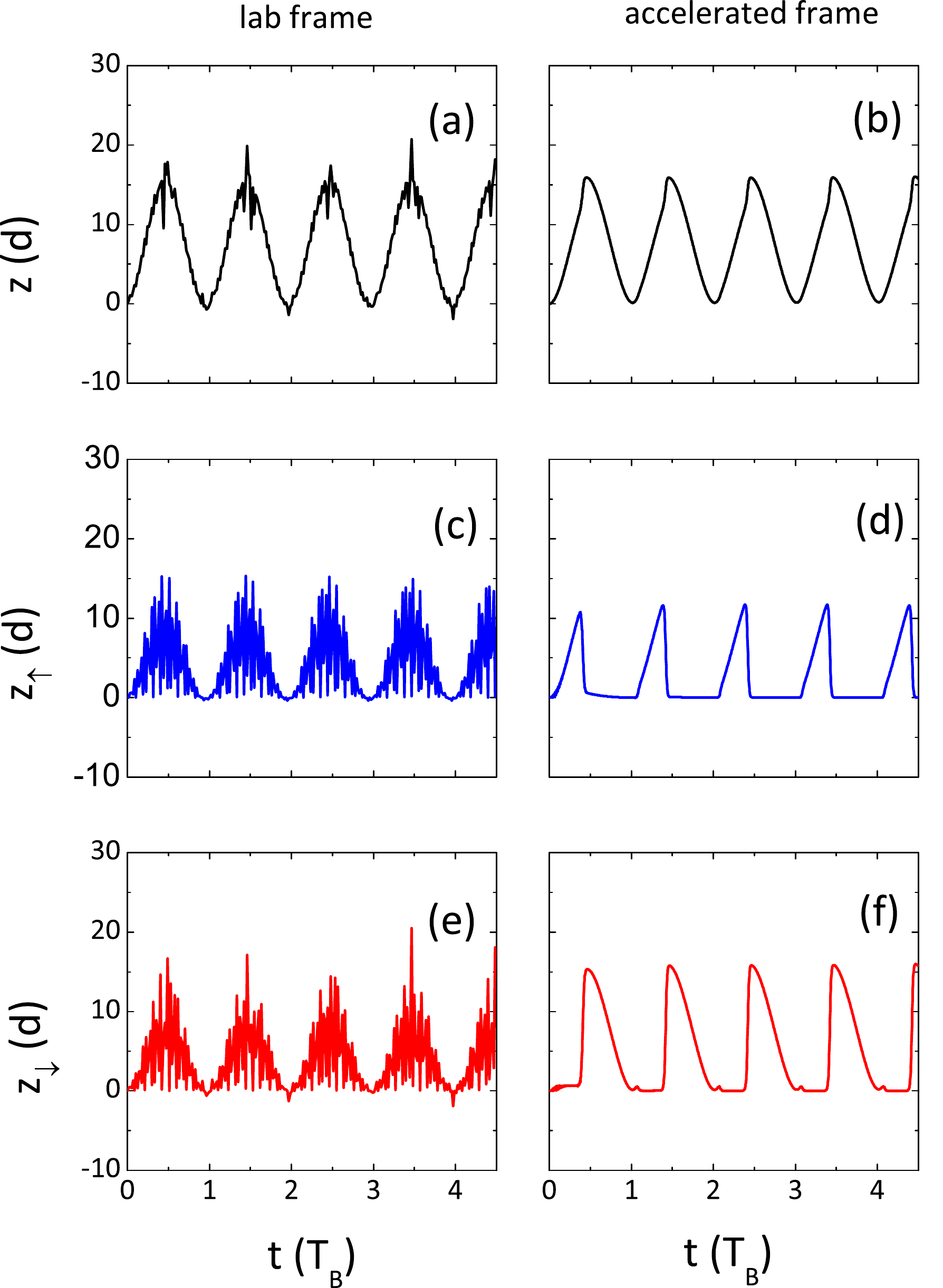}\caption{{\protect\footnotesize (Color
online) Oscillation dynamics in the lab frame (left column) and the
accelerated frame (right column). The dynamics of }$z=z_{\uparrow
}+z_{\downarrow}$ {\protect\footnotesize (first row), }$z_{\uparrow}$
{\protect\footnotesize (middle row) and }$z_{\downarrow}$
{\protect\footnotesize (bottom row) are shown in black, red and blue lines
respectively. The parameters are set as }$\gamma=0.2${\protect\footnotesize ,
}$J=10Fd${\protect\footnotesize , }$\delta=0${\protect\footnotesize , }%
$\hbar\Omega=80Fd$ {\protect\footnotesize and }$E_{r}/J=8.55$%
{\protect\footnotesize .}}%
\label{fig_galilean}%
\end{figure}

In the case without SO coupling, one can introduce a linearly time-dependent
frequency difference $\Delta\nu\left(  t\right)  =-Ft/md$ between the two
lattice beams \cite{castin1996}, the lattice potential becomes $U_{0}\sin
^{2}\left[  k_{l}z-\pi\int_{0}^{t}d\tau\Delta\nu\left(  \tau\right)  \right]
$ and in an \textit{accelerated} frame it is equivalent to exerting a constant
inertial force $F$ on the atoms trapped in a stationary lattice. However this
equivalence cannot be established in the presence of SO coupling. This is due
to that the SO Hamiltonian $\hat{H}_{SO}$ breaks Galilean invariance as the
physical momentum $p_{z}-\hat{A}$ does not commute with $\hat{H}_{SO}$. In
this case going into a moving inertial frame will result in an additional
time-dependent term $-\alpha Ft\hat{\sigma}_{z}$ in Hamiltonian
(\ref{eq_single_particle_Hamiltonian}), which play the role as a
time-dependent effective detuning.

We calculate the oscillation dynamics in the stationary frame (lab frame) with
the exerting force $F$ and that in the accelerated frame within which the
atoms are subject to an effective force $F$ as well as an effective
time-dependent detuning $-\alpha Ft\hat{\sigma}_{z}$, the results are shown in
Fig. \ref{fig_galilean}. The dynamics in the lab frame are simulated with
eigenstate expansion while that in the accelerated frame are calculated by
means of the Fourth-order Runge-Kutta method. Both the initial state are given
by Eq. (\ref{eq_initial}). In the numerical simulation we consider the recoil
energy $E_{r}=\hbar^{2}k_{l}^{2}/2m=8.55J$ for a typical experimental value of
$U_{0}=4E_{r}$. As one can expect, in the lab frame the oscillation dynamics
for spin-$\uparrow$ and $\downarrow$ components are in phase, as shown in
Figs. \ref{fig_galilean}(c) and (e). However the dynamics shown in Figs.
\ref{fig_galilean}(d) and (f) indicate that they are out-of-phase (phase
separated in the time domain) in the accelerated frame. This interesting
dynamics can be readily captured in experiment and serve as a clear proof of
broken Galilean invariance, which is also the mechanism underlying other
unusual behaviors such as the deviation of dipole oscillation frequency in a
harmonically trapped system \cite{chenPRL2012,puPRA2012}, the ambiguity in
defining Landau critical velocity in SO coupled condensates \cite{wuEPL2012},
finite-momentum dimer bound state in a SO coupled Fermi gas \cite{dongPRA2013}
and asymmetric expansion of SO coupled atomic Bose gas \cite{engelsPRL2017}.
The effect of broken Galilean invariance can be signified via introducing a
frequency difference between the two laser beams forming the optical lattice
\cite{EngelsPRL2015}.

\section{spin current generation}

\label{sec_current}

An interesting question is how to create a spin current with SO coupling
\cite{SpinCurrentScience}. Spin current have been experimentally generated in
a SO-coupled BEC via spin Hall effect \cite{spielmanNature2013} and quenching
\cite{SpinCurrentQuench}. In theory, Larson \textit{et al}. studied bloch
oscillations of atomic BEC in a tilted two-dimensional (2D) optical lattice
\cite{LarsonPRA2010}, in which the atoms are subject to a 2D SO interaction
$\hat{H}_{SO}\propto\hat{p}_{x}\hat{\sigma}_{x}+\hat{p}_{y}\hat{\sigma}_{y}$
and in turn give rise to a spin-dependent effective force propotional to
$\hat{\sigma}_{z}$. As a result an oscillating transverse spin current can be
generated. For the present 1D system we have%
\begin{equation}
\hat{F}_{z}=\left[  \hat{H}_{SO},\left[  \mathbf{\hat{z}},\hat{H}_{SO}\right]
\right]  =\frac{\hbar^{3}k_{R}\Omega}{m}\hat{\sigma}_{y}\mathbf{e}_{z},
\label{eq_force}%
\end{equation}
indicating an SO aroused effective force along $\mathbf{e}_{z}$-direction and
propotional to $\hat{\sigma}_{y}$.

\ Here we would like to explore the possibility of generating spin current in
the present 1D system with this effective force. As suggested by Shi
\textit{et al}. \cite{spin current}, the spin current operator along the
$z$-direction can be defined as%
\begin{equation}
\hat{J}_{S}^{i}\left(  t\right)  =\frac{d}{dt}\left(  \hat{\sigma}_{i}\hat
{z}\right)  .\label{eq_spin current}%
\end{equation}
Follow the very similar procedure as deducing Eqs. (\ref{eq_velocity1}) and
(\ref{eq_velocity2}), make use of the WSL eigenstate, the mean-value of
$\sigma_{z}$-component of spin current can be calculated as%
\begin{align}
J_{S}^{z}\left(  t\right)   &  =\sum_{j}\operatorname{Im}\left[  \frac
{Jd}{\hbar}e^{-i\pi\gamma}\psi_{j,\uparrow}^{\ast}\left(  t\right)
\psi_{j+1,\uparrow}\left(  t\right)  \right.  \nonumber\\
&  \left.  -\frac{Jd}{\hbar}e^{i\pi\gamma}\psi_{j,\downarrow}^{\ast}\left(
t\right)  \psi_{j+1,\downarrow}\left(  t\right)  +2\Omega dj\psi_{j,\uparrow
}^{\ast}\left(  t\right)  \psi_{j,\downarrow}\left(  t\right)  \right]
\nonumber\\
&  =\sum_{\nu,\nu^{\prime}}\operatorname{Im}\left\{  a_{\nu}^{\ast}%
a_{\nu^{\prime}}\sum_{j}\left[  \frac{Jd}{\hbar}W_{j,\uparrow}^{\nu\ast
}W_{j+1,\uparrow}^{\nu^{\prime}}e^{-i\pi\gamma}\right.  \right.  \nonumber\\
&  \left.  \left.  -\frac{Jd}{\hbar}W_{j,\downarrow}^{\nu\ast}%
W_{j+1,\downarrow}^{\nu^{\prime}}e^{i\pi\gamma}+2\Omega djW_{j,\uparrow}%
^{\nu\ast}W_{j,\downarrow}^{\nu^{\prime}}\right]  e^{i\left(  E_{\nu}%
-E_{\nu^{\prime}}\right)  t/\hbar}\right\}  ,\label{eq_current1}%
\end{align}
which predicts that in addition to the coupling between different rungs, the
last term in Eq. (\ref{eq_current1}) indicate that the coupling between
spin-$\uparrow$ and $\downarrow$ components also contribute to the spin
current, resulting from that the effective force $\hat{F}_{z}$ is propotional
to $\hat{\sigma}_{y}$.\begin{figure}[th]
\includegraphics[width=8cm]{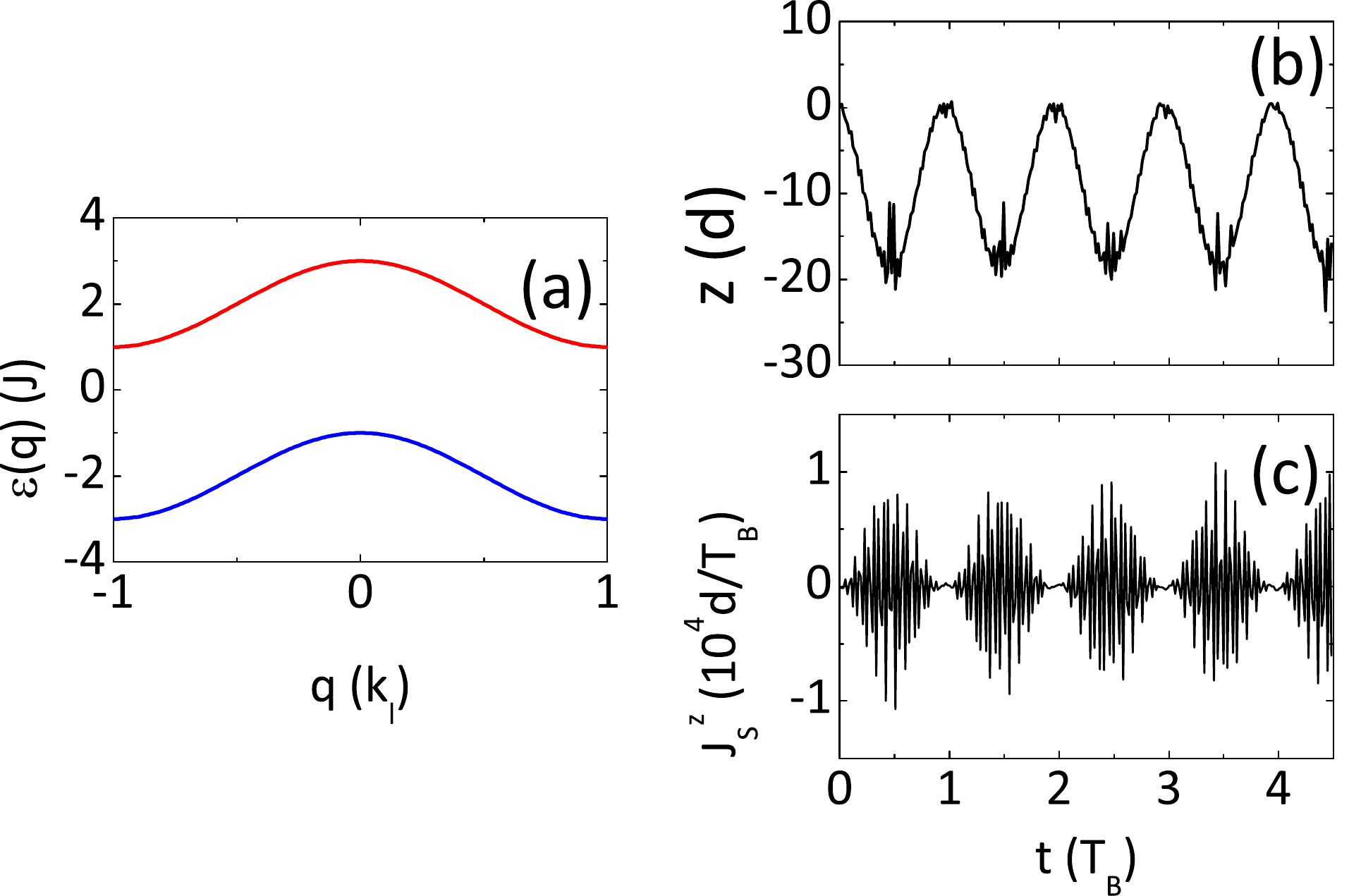}\caption{{\protect\footnotesize (Color online)
(a) Energy band at }$\gamma=1$ {\protect\footnotesize and }$\delta
=0${\protect\footnotesize . (b) Dynamics of mean position }$z$%
{\protect\footnotesize . (c) Dynamics of the spin current }$J_{S}^{z}%
${\protect\footnotesize .\ The parameters are set as }$J=10Fd$
{\protect\footnotesize and }$\hbar\Omega=80Fd$ {\protect\footnotesize with the
atoms initially prepared in a Gaussian wavepacket (\ref{eq_initial}).}}%
\label{fig_spin current}%
\end{figure}

In order to illustrate the contribution of this term, one can choose
$\gamma=1$ at which the major intraladder contribution from first two terms in
Eq. (\ref{eq_current1}) canceled out at $\delta=0$ due to the symmetry.
Physically it is equivalent to that the two spin components are performing
identical Bloch oscillation and in the meanwhile subject to on-site Raman
coupling, as one can see from the Hamiltonian
(\ref{eq_single_band_Hamiltonian}). In the case $\left\langle \hat{\sigma}%
_{z}\right\rangle =0$ for the Bloch eigenstate without the force. We then
numerically calculate $J_{S}^{z}\left(  t\right)  $ and the results are shown
in Fig. \ref{fig_spin current}. One can expect that in the absence of Raman
coupling no spin current can be generated since that spin-$\uparrow$ and
spin-$\downarrow$ components both move adiabatically along the energy band and
exhibit typical properties of Bloch oscillation, as can be seen from Fig.
\ref{fig_spin current}(a) and (b). The small amplitude high frequency
oscillation is aroused by interladder coupling as we discussed in Sec.
\ref{sec_bloch}. The time evolution of spin current $J_{S}^{z}$ exhibit the
behavior of collapse and revival shown in Fig. \ref{fig_spin current}(c),
reminiscent of the Jaynes-Cummings model in quantum optics \cite{JC
experiment}. This collapse and revival behavior can be understood as a result
of the complex interplay between the external force $F$ and the intrinsic
force $F_{SO}=\hbar k_{R}\Omega\hat{\sigma}_{y}/2$ aroused by SO interaction.
One can also understand this collapse and revival behavior the same as
Zitterbewegung \cite{LarsonPRA2010}. Zitterbewegung results from coherent
coupling between eigenstates of Dirac cone with different helicity
\cite{Zitterbewegung} and have been successfully observed in experiment with
cold atoms \cite{leblancNJP2013,quPRA2013R}, while here the trembling
oscillation is aroused by spin swapping.

We also examined the case with finite Zeeman detuning. As one can expect,
although the spin-$\uparrow$ and spin-$\downarrow$ components are performing
different oscillation, it will be immersed in the dynamics aroused by Raman
coupling and in general the spin current will exhibit the dynamics of collapse
and revival. In order to achieve constant directional spin current, one can
either adapt time-dependent SO coupling \cite{chienPRA2013} or unbiased
external force \cite{spin ratchet}.

\section{summary and outlook}

\label{sec_summary}

Before concluding the paper, we need to note that in the presence of SO
interaction one should be very careful while using the above lowest energy
band truncation. As was pointed out by Zhou and Cui \cite{cuiPRB2015}, in this
case tight-binding models have limitations in predicting the correct
single-particle physics due to the missed high-band contributions. Physically
the Raman lasers inducing SO interaction also inevitably couple atoms to
high-lying bands which will significantly affect the single-particle physics
\cite{panPRA2016}. Experimentally atomic BEC can also be prepared in excited
bands of an optical lattice \cite{zhouxjPRA2013}. Ao and Rammer also pointed
out that high-band contributions can substantially affect the Bloch
oscillation dynamics \cite{pingao}. Contributions from higher Bloch bands will
be important and interesting in orbital optical lattices
\cite{hemmerichreview}. By considering that, we compare the results presented
in this work with those obtained through numerical simulation of the
corresponding Schrodinger equation and found good agreement in the case of
large energy gap and small external force.

In summary, we have studied the Bloch oscillation dynamics of a SO-coupled
cold atomic gas trapped inside a 1D optical lattice. The eigen-spectra of the
system have been identified as two interpenetrating WSL. The Bloch oscillation
dynamics in this system can be well-understood via analyzing the coupling
between different rungs of the WSL. In the presence of finite Zeeman detuning,
we show that the two spin components can display out-of-phase oscillation.
This can also serve as an unambiguous proof of broken Galilean invariance
aroused by SO coupling. In addition to that, we numerically explored the
possibility of generating spin current in the present system. Since SO
interaction have been implemented in BEC in a 1D optical lattice
\cite{EngelsPRL2015}, our findings of the interesting dynamical phenomena
should be within reach of present-day experiments. For BEC it will be
interesting to study the impact of interparticle collisions on Bloch
oscillation \cite{gaulPRA2011} and spin current generation, which can be
investigated by the Gaussian variational approach \cite{smerzi,chenPRA2014}.
It will also be interesting to investigate Landau-Zener tunneling
\cite{LeePRA2015,wuNPJ2003}. These will be left for further investigation.

\begin{acknowledgments}
We thank Han Pu and Yongping Zhang for careful reading and many helpful
comments on the manuscript. This work is supported by the National Natural
Science Foundation of China (Grants No. 11374003, No. 11774093, No. 11574086,
No. 91436211, No. 11654005), the National Key Research and Development Program
of China (Grant No. 2016YFA0302001) the Shanghai Rising-Star Program (Grant
No. 16QA1401600), and the Science and Technology Commission of Shanghai
Municipality (Grants No. 16DZ2260200 and No. 16ZR1409800).
\end{acknowledgments}


\begin{thebibliography}{99}                                                                                               %


\bibitem {blochoriginal}F. Bloch, Z. Phys. \textbf{52}, 555 (1929); C. Zener,
Proc. R. Soc. London A \textbf{145}, 523 (1934).

\bibitem {blochexperiment}C. Waschke, H. G. Roskos, R. Schwedler, K. Leo, H.
Kurz, and K. K\"{o}hler, Phys. Rev. Lett \textbf{70}, 3319 (1993).

\bibitem {topology}L.-K. Lim, J.-N. Fuchs, and G. Montambaux, Phys. Rev. Lett.
\textbf{108}, 175303 (2012); Y.-Q. Wang and X.-J. Liu, Phys. Rev. A
\textbf{94}, 031603(R) (2016).

\bibitem {opticalBloch}T. Pertsch, P. Dannberg, W. Elflein, A. Br\"{a}uer, and
F. Lederer, Phys. Rev. Lett \textbf{83}, 4752 (1999); Y. Zhang, D. Zhang, Z.
Zhang, C. Li, Y. Zhang, F. Li, M. R. Beli\'{c}, and M. Xiao, Optica
\textbf{4}, 571 (2017).

\bibitem {castin1996}M. B. Dahan, E. Peik, J. Reichel, Y. Castin, and C.
Salomon, Phys. Rev. Lett. \textbf{76}, 4508 (1996); E. Peik, M. B. Dahan, I.
Bouchoule, Y. Castin, and C. Salomon, Phys. Rev. A \textbf{55}, 2989 (1997).

\bibitem {niuPRL1996}Q. Niu, X.-G. Zhao, G. A. Georgakis, and M. G. Raizen,
Phys. Rev. Lett \textbf{76}, 4504 (1996); S. R. Wilkinson, C. F. Bharucha, K.
W. Madison, Q. Niu, and M. G. Raizen, Phys. Rev. Lett \textbf{76}, 4512 (1996).

\bibitem {atomBlochreview}For a review, see M. G. Raizen, C. Salomon, and Q.
Niu, Phys. Today \textbf{50}, 30 (1997), and references therein.

\bibitem {impurity}F. Meinert, M. Knap, E. Kirilov, K. Jag-Lauber, M. B.
Zvonarev, E. Demler, and H.-C. N\"{a}gerl, Science \textbf{356}, 945 (2017).

\bibitem {impuritytheory}D. M. Gangardt and A. Kamenev, Phys. Rev. Lett
\textbf{102}, 070402 (2009).

\bibitem {WSLreview}For a review, see M. Gluck, A. R. Kolovsky, and H. J.
Korsch, Phys. Rep. \textbf{366}, 103 (2002), and references therein.

\bibitem {chenPRL2012}J.-Y. Zhang, S.-C. Ji, Z. Chen, L. Zhang, Z.-D. Du, B.
Yan, G.-S. Pan, B. Zhao, Y.-J. Deng, H. Zhai, S. Chen, and J.-W. Pan, Phys.
Rev. Lett. \textbf{109}, 115301 (2012).

\bibitem {1dso}Y.-J. Lin, K. Jim\'{e}nez-Garcia and I. B. Spielman, Nature
\textbf{471}, 83 (2011). P. Wang, Z.-Q. Yu, Z. Fu, J. Miao, L. Huang, S. Chai,
H. Zhai, and J. Zhang, Phys. Rev. Lett. \textbf{109}, 095301 (2012). L. W.
Cheuk, A. T. Sommer, Z. Hadzibabic, T. Yefsah, W. S. Bakr, and M. W.
Zwierlein, Phys. Rev. Lett. \textbf{109}, 095302 (2012).

\bibitem {soReview}For a review, see J. Dalibard, F. Gerbier, G. Juzeliunas,
and P. \"{O}hberg, Rev. Mod. Phys. \textbf{83}, 1523 (2011); N. Goldman, G.
Juzeli\={u}nas, P. \"{O}hberg, I. B. Spielman, Rep. Prog. Phys. \textbf{77},
126401 (2014); H. Zhai, Rep. Prog. Phys. \textbf{78}, 026001 (2015).

\bibitem {puPRA2012}B. Ramachandhran, B. Opanchuk, X.-J. Liu, H. Pu, P. D.
Drummond, and H. Hu, Phys. Rev. A \textbf{85}, 023606 (2012).

\bibitem {leblancNJP2013}L. J. LeBlanc, M. C. Beeler, K. J.-Garcia, A. R.
Perry, S. Sugawa, R. A. Williams, and I. B. Spielman, New J. Phys.
\textbf{15}, 073011 (2013).

\bibitem {quPRA2013R}C. Qu, C. Hamner, M. Gong, C. Zhang, and P. Engels, Phys.
Rev. A \textbf{88}, 021604(R) (2013).

\bibitem {dongPRA2013}L. Dong, L. Jiang, H. Hu, and H. Pu, Phys. Rev. A
\textbf{87}, 043616 (2013).

\bibitem {zhouPRA2013}L. Zhou, H. Pu, and W. Zhang, Phys. Rev. A \textbf{87},
023625 (2013).

\bibitem {orsoPRL2017}G. Orso, Phys. Rev. Lett \textbf{118}, 105301 (2017).

\bibitem {shermanPRL2015}S. Mardonov, M. Modugno, and E. Y. Sherman, Phys.
Rev. Lett \textbf{115}, 180402 (2015).

\bibitem {zhouatomoptics}L. Zhou, J.-L. Qin, Z. Lan, G. Dong, and W. Zhang,
Phys. Rev. A \textbf{91}, 031603(R) (2015); L. Zhou, R.-F. Zheng, and W.
Zhang, Phys. Rev. A \textbf{94}, 053630 (2016).

\bibitem {EngelsPRL2015}C. Hamner, Y. Zhang, M.\thinspace A. Khamehchi, M. J.
Davis, and P. Engels, Phys. Rev. Lett. \textbf{114}, 070401 (2015).

\bibitem {flat band}Y. Zhang and C. Zhang, Phys. Rev. A \textbf{87}, 023611 (2013).

\bibitem {panPRA2016}J.-S. Pan, W. Zhang, W. Yi, and G.-C. Guo, Phys. Rev. A
\textbf{94}, 043619 (2016).

\bibitem {engelsPRL2017}M.\thinspace A. Khamehchi, K. Hossain, M.\thinspace E.
Mossman, Y. Zhang, T. Busch, M. M. Forbes, and P. Engels, Phys. Rev. Lett
\textbf{118}, 155301 (2017).

\bibitem {lanPRA2014}Z. Lan and P. \"{O}hberg, Phys. Rev. A \textbf{89},
023630 (2014).

\bibitem {LarsonPRA2010}J. Larson, J.-P. Martikainen, A. Collin, and E.
Sj\"{o}qvist, Phys. Rev. A \textbf{82}, 043620 (2010).

\bibitem {caetanoPRB2014}R. A. Caetano, Phys. Rev. B \textbf{89}, 195414 (2014).

\bibitem {KartashovPRL2016}Y. V. Kartashov, V. V. Konotop, D. A. Zezyulin, and
L. Torner, Phys. Rev. Lett. \textbf{117}, 215301 (2016).

\bibitem {SpinCurrentScience}P. Sharma, Science \textbf{307}, 531 (2005).

\bibitem {peierls}R. E. Peierls, Z. Phys. \textbf{80}, 763 (1933); D. R.
Hofstadter, Phys. Rev. B \textbf{14}, 2239 (1976).

\bibitem {lattice parameter}A. A. Mostofi, J. R. Yates, Y.-S. Lee, I. Souza,
D. Vanderbilt, and N. Marzari, Comput. Phys. Commun. \textbf{178}, 685 (2008);
R. Walters, G. Cotugno, T. H. Johnson, S. R. Clark, and D. Jaksch, Phys. Rev.
A \textbf{87}, 043613 (2013).

\bibitem {HartmannNPJ2004}T. Hartmann, F. Keck, H. J. Korsch, and S. Mossmann,
New J. Phys. \textbf{6}, 2 (2004).

\bibitem {Wannier1960}G. H. Wannier, Phys. Rev. \textbf{117}, 432 (1960).

\bibitem {symmetry}V. E. Lobanov, Y. V. Kartashov, and V. V. Konotop, Phys.
Rev. Lett. \textbf{112}, 180403 (2014); Y. V. Kartashov, V. V. Konotop, and D.
A. Zezyulin, Europhys. Lett. \textbf{107}, 50002 (2014).

\bibitem {wuEPL2012}Q. Zhu, C. Zhang, and B. Wu, Europhys. Lett. \textbf{100},
50003 (2012).

\bibitem {spielmanNature2013}M. C. Beeler, R. A. Williams, K.
Jim\'{e}nez-Garc\'{\i}a, L. J. LeBlanc, A. R. Perry and I. B. Spielman, Nature
\textbf{498}, 201 (2013).

\bibitem {SpinCurrentQuench}C.-H. Li, C. Qu, R. J. Niffenegger, S.-J. Wang, M.
He, D. B. Blasing, A. Olson, C. H. Greene, Y. Lyanda-Geller, Q. Zhou, C. Zhang
and Y. P. Chen, arXiv:1810.06504.

\bibitem {spin current}J. Shi, P. Zhang, D. Xiao, and Q. Niu, Phys. Rev. Lett.
\textbf{96}, 076604 (2006).

\bibitem {JC experiment}G. Rempe, H. Walther, and N. Klein, Phys. Rev. Lett.
\textbf{58}, 353 (1987).

\bibitem {Zitterbewegung}E. Schr\"{o}dinger, Sitzungsber. Preuss. Akad. Wiss.
Phys. Math. Kl. \textbf{24}, 418 (1930).

\bibitem {chienPRA2013}C.-C. Chien and M. Di Ventra, Phys. Rev. A \textbf{87},
023609 (2013).

\bibitem {spin ratchet}S. Smirnov, D. Bercioux, M. Grifoni, and K. Richter,
Phys. Rev. B \textbf{78}, 245323 (2008).

\bibitem {cuiPRB2015}L. Zhou and X. Cui, Phys. Rev. B \textbf{92}, 140502(R) (2015).

\bibitem {zhouxjPRA2013}Y. Zhai, X. Yue, Y. Wu, X. Chen, P. Zhang, and X.
Zhou, Phys. Rev. A \textbf{87}, 063638 (2013).

\bibitem {pingao}P. Ao, Phys. Rev. B \textbf{41}, 3998 (1990); P. Ao and J.
Rammer, Phys. Rev. B \textbf{44}, 11494 (1991).

\bibitem {hemmerichreview}For a review, see T. Kock, C. Hippler, A. Ewerbeck,
and A. Hemmerich, J. Phys. B \textbf{49}, 042001 (2016), and references therein.

\bibitem {gaulPRA2011}C. Gaul, E. D\'{\i}az, R. P. A. Lima, F.
Dom\'{\i}nguez-Adame, and C. A. M\"{u}ller, Phys. Rev. A \textbf{84}, 053627 (2011).

\bibitem {smerzi}A. Trombettoni and A. Smerzi, Phys. Rev. Lett. \textbf{86},
2353 (2001).

\bibitem {chenPRA2014}Y. Cheng, G. Tang, and S. K. Adhikari, Phys. Rev. A
\textbf{89}, 063602 (2014).

\bibitem {wuNPJ2003}B. Wu and Q. Niu, New J. Phys. \textbf{5}, 104 (2003).

\bibitem {LeePRA2015}Y. Ke, X. Qin, H. Zhong, J. Huang, C. He, and C. Lee,
Phys. Rev. A \textbf{91}, 053409 (2015).
\end{thebibliography}
\end{document}